\documentclass[english,aps,twocolumn]{revtex4}
\usepackage[latin1]{inputenc}
\usepackage{graphicx}

\makeatletter

\providecommand{\LyX}{L\kern-.1667em\lower.25em\hbox{Y}\kern-.125emX\@}
\usepackage{babel}
\makeatother
\begin{document}

\preprint{This line only printed with preprint option}

\title{Experimental Demonstration of Macroscopic Quantum Coherence in Gaussian States}

\author{Christoph Marquardt$^1$}
\email{marquardt@kerr.physik.uni-erlangen.de}
\author{Ulrik L. Andersen$^{1,2}$}
\author{Gerd Leuchs$^1$}
\affiliation{$^1$Institut f\"{u}r Optik, Information und Photonik, Max-Planck Forschungsgruppe, Universit\"{a}t Erlangen-N\"{u}rnberg, G\"{u}nther-Scharowsky-Str. 1, 91058, Erlangen, Germany \\
$^2$Department of Physics,
Technical University of Denmark,
Building 309, 2800 Lyngby, Denmark}

\author{Yuishi Takeno}
\author{Mitsuyoshi Yukawa}
\author{Hidehiro Yonezawa}
\author{Akira Furusawa}
\affiliation{Department of Applied Physics, School of Engineering, The University of Tokyo, 7-3-1 Hongo, Bunkyo-ku, Tokyo 113-8656, Japan}

\begin{abstract}
We witness experimentally the presence of macroscopic coherence in Gaussian quantum states using a recently proposed criterion (E.G. Cavalcanti and M. Reid, Phys. Rev. Lett. {\bf 97} 170405 (2006)). The macroscopic coherence stems from interference between macroscopically distinct states in phase space and we prove experimentally that even the vacuum state contains these features with a distance in phase space of $0.51\pm0.02$ shot noise units (SNU). For squeezed states we found macroscopic superpositions with a distance of up to $0.83\pm0.02$ SNU. The proof of macroscopic quantum coherence was investigated with respect to squeezing and purity of the states.
\end{abstract}
\maketitle

%\section{introduction}

Quantum mechanics has led to many peculiar effects that were not easily
conceivable with everyday perception. Famous examples are the concept of
quantized energy~\cite{quantization}, the double-slit
experiment~\cite{doubleslit} and the Einstein-Podolsky-Rosen (EPR) gedanken experiment~\cite{EPR}. Another striking example was introduced by Schr\"odinger in 1935, when discussing quantum
superpositions of macroscopically distinct states. In his famous
 gedanken experiment a cat could be in a state of being neither dead nor alive~\cite{scat}. By assigning the two quantum states `dead' and `alive' by $\mid\Psi_{+}\rangle$ and $\mid\Psi_{-}\rangle$, the envisioned state of the cat is the coherent superposition $\mid\Psi_{+}\rangle+\mid\Psi_{-}\rangle$ until it is observed and collapses into one of the two macroscopically distinct states, thus determining the fate of the cat. Microscopic superposition states such as in two-level atoms are readily accepted, whereas the macroscopic superposition state such as the Schr\"odinger cat state is considered counter-intuitive and hardly imaginable. However, in recent years there have been several attempts to produce superposition states approaching the macroscopic regime~\cite{experimentsscat}. A severe hindrance for the production of these states, however, is decoherence associated with the unavoidable coupling to the surrounding reservoir which causes the system to evolve into a classical mixture~\cite{decoherence}.

Recently Cavalcanti and Reid introduced the concept of generalized
macroscopic superpositions~\cite{Cavalcanti}. Instead of the original
example of $\mid\Psi_{+}\rangle+\mid\Psi_{-}\rangle$ with two macroscopically distinct states, the
generalized state is a three component coherent superposition of the form $\mid\Psi_{+}\rangle + \mid\Psi_{0}\rangle + \mid\Psi_{-}\rangle$. Thus in addition to the two macroscopically distinct states, an 'intermediate' state, $|\Psi_{0}\rangle$, was introduced. The neighboring pairs of states, that is ($\mid\Psi_{+}\rangle,\mid\Psi_{0}\rangle$) and ($\mid\Psi_{0}\rangle,\mid\Psi_{-}\rangle$), might be microscopically distinct as witnessed by non zero off diagonal density matrix elements close to the diagonal. It is, however, still possible to have the two outer states $\mid\Psi_{+}\rangle$ and $\mid\Psi_{-}\rangle$ macroscopically distinct. This macroscopic coherence, hidden in the overall microscopic state, is reflected by the most off-diagonal element ($\langle\Psi_{+}\mid\rho\mid\Psi_{-}\rangle$) in the density matrix ($\rho$) being non-zero. As shown by Cavalcanti and Reid, such macroscopic coherences appear in various common states and can be witnessed through simple homodyne measurements of conjugate variables~\cite{Cavalcanti}.    

In this Letter we experimentally witness  the presence of macroscopic coherence in various states employing the criteria put forward in ref.~\cite{Cavalcanti}. The states under interrogations are vacuum, squeezed and entangled states, all of which are proven to contain macroscopic coherence to some extent. We also investigate the sensitivity of the macroscopicality with regard to the degree of squeezing and purity of the squeezed states.

\begin{figure}
\includegraphics[scale=0.6]{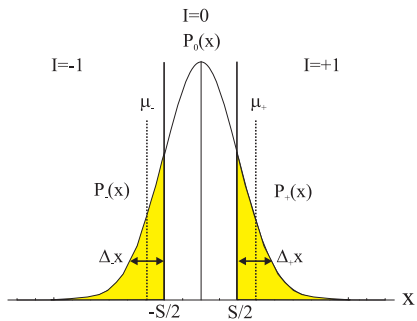}

\caption{\label{fig:concept}P(x) is one of the marginal distributions of the Wigner function. Binning of one quadrature variable into negative,
intermediate and positive regions, with intermediate distance $S$.}
\end{figure}

We start by shortly reviewing the definition of a generalized superposition state as presented in
ref. \cite{Cavalcanti}. This state is given by
\begin{equation}
\mid\Phi\rangle=c_{+}\mid\Psi_{+}\rangle+c_{0}\mid\Psi_{0}\rangle+c_{-}\mid\Psi_{-}\rangle,\label{eq:reidmacrosup1}\end{equation}
with the probability amplitudes $c_{+},c_{-}\neq0$. Measuring the state $|\Phi\rangle$ with the projectors $\mid\Psi_{+}\rangle\langle\Psi_{+}|$,$\mid\Psi_{0}\rangle\langle\Psi_{0}|$ and $\mid\Psi_{-}\rangle\langle\Psi_{-}|$ results in the outcomes $+1$,$-1$ and $0$ (with probabilities $|c_+|^2$,$|c_-|^2$ and $|c_0|^2$) where the +1 and -1 outcomes are macro- or mesoscopically distinguishable. 
In the case of quadrature measurements,
a possible separation of the outcomes to yield an appropriate projector is illustrated in Fig.~\ref{fig:concept}.
The results of the measurement of a quadrature variable $x$ is divided
into three distinct regions $I=-1,0,+1$ corresponding to the outcomes
of the states mentioned (with the probabilities $P_{-}$, $P_{0}$ and $P_{+}$ to get a result in that region). The $I=-1$ and $I=+1$ regions are separated by a
distance $S$ giving a measure for the macroscopicality of the generalized
superpositions. 

The next question is how to measure the existence of these superpositions.
It is neither realistic nor feasible to construct a measurement device that projects directly onto the superposition state $c_{+}\mid\Psi_{+}\rangle+c_{-}\mid\Psi_{-}\rangle$, such an apparatus would be highly complex and the projected state necessarily highly sensitive to decoherence~\cite{pointerstates}. Alternatively, the presence of macroscopic coherence can be witnessed by tomographic reconstruction of the state's density matrix in the basis spanned by the eigenstates, $\mid\Psi_{+}\rangle$, $\mid\Psi_{0}\rangle$ and $\mid\Psi_{-}\rangle$. The non zero values of the relevant off-diagonal elements witness the superposition. A much simpler approach to prove the existence of macroscopic coherence was developed by Cavalcanti and Reid~\cite{Cavalcanti}. They showed that by applying a certain criterion, the presence of a macroscopic superposition state can be verified through simple ensemble measurements of conjugate quadratures. 

\begin{figure}
\includegraphics[scale=0.25]{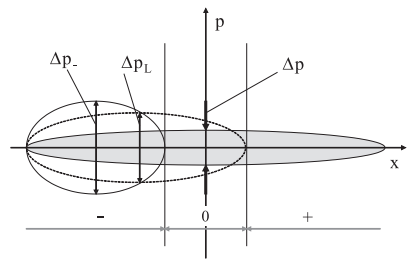}

\caption{\label{fig:delta-p_s-in-phase-diagram}According to Heisenberg's uncertainty relation the minimum variance of $p$ increases when restricting the distribution of $x$-values to one subspace. The smaller variance of $p$ in full space is evidence for the overall state being a coherent superposition of the subspace states.}
\end{figure}

We will now sketch the basic idea used in~\cite{Cavalcanti}. If phase space is divided into the three subspaces as indicated in Fig.~\ref{fig:concept} and no correlations between the subspaces are assumed, the overall variance of the $p$-variable in the mixed state is the weighted sum of the variances of $p$ in the individual regions:
\begin{equation}
\Delta_{mixed}^{2}p\geq P_{-}\Delta_{-}^{2}p + P_{0}\Delta_{0}^{2}p + P_{+}\Delta_{+}^{2}p. \label{eq:reid-unequality-rudimentary}
\end{equation}
Without coherences between subspaces Heisenberg's uncertainty relation applies to each subspace separately and $\Delta_{i}^{2}p\geq\Delta^{2}p\,$ for $i=-,0,+$ (see Fig.~\ref{fig:delta-p_s-in-phase-diagram}).
A violation of inequality (\ref{eq:reid-unequality-rudimentary}) is evidence for coherences between the different subspaces. At this point microscopic superpositions would suffice to violate the inequality. Therefore, Cavalcanti and Reid~\cite{Cavalcanti} replaced $ \Delta_{-}^{2}p$ by the smaller variance $\Delta_{L}^{2}p$ etc. which takes any microscopic superpositions between neighboring regions into account. The resulting new inequality is now only violated if there are macroscopic superpositions between the two outer well separated subspaces. The final step is expressing the various variances of $p$ by quantities which are straight forward to measure. The resulting inequality   

\begin{equation}
(\Delta_{ave}^{2}x+P_{0}\delta)\Delta^{2}p\geq1.\label{eq:reid-unequality}
\end{equation}
is sufficient to prove the presence of generalized superpositions of
the form (\ref{eq:reidmacrosup1}) with a distance S in phase space.     
The variance $\Delta_{ave}^{2}x$ is defined as $\Delta_{ave}^{2}x=P_{+}\Delta_{+}^{2}x+P_{-}\Delta_{-}^{2}x,$
with $\Delta_{+}^{2}x$ and $\Delta_{-}^{2}x$ being the variances of the distributions associated with the 
regions $I=+1$ and $I=-1$ (see Fig.~\ref{fig:concept}), and
$\Delta^{2}p$ is the variance of the conjugate variable $p$. The
distance S contributes to
\begin{equation}
\delta=(\mu_{+}+S/2)^{2}+(\mu_{-}-S/2)^{2}+S^{2}/2+\Delta_{+}^{2}x+\Delta_{-}^{2}x,\label{eq:reid_delta}\end{equation}
where $\mu_{+}$ and $\mu_{-}$ are the mean values of the distributions associated with the regions
$I=+1$ and $I=-1$.

\begin{figure}
\includegraphics[scale=0.6]{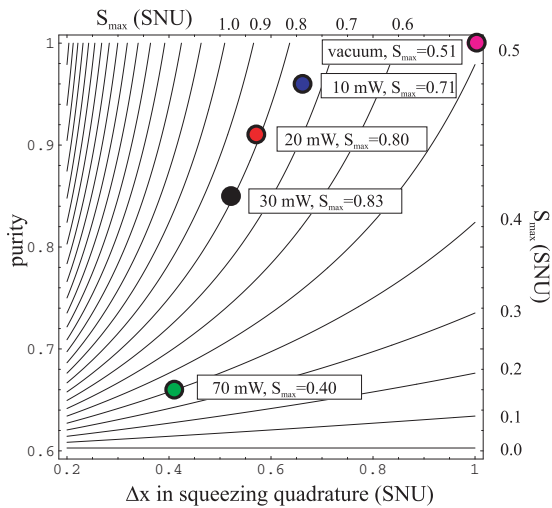}

\caption{\label{fig:smaxsqpurity} Contour plot of the maximum distance $S_{max}$, that can be proven with a
squeezed state of given purity and squeezing. Symbols denote the measured values.}
\end{figure}

Inequality (\ref{eq:reid-unequality}) determines the maximum distance $S$, for which a generalized
superposition can be proven for squeezed Gaussian
states. $S_{max}$ depends on the degree of squeezing as well as on the purity
of the squeezed states. For pure squeezed states generalized superpositions exist 
for an $S_{max}$ of about 0.5 of the standard deviation of the marginal probability distribution. Therefore,
by squeezing the $p$ quadrature, the associated anti-squeezing of $x$ enables the violation for larger
distances $S$, eventually reaching a truly macroscopic regime for large degrees of squeezing. However, in practice, the production of highly squeezed states is often accompanied with decoherence, which on the other hand makes it harder to violate the inequality (\ref{eq:reid-unequality}) with large $S$. Hence there exists a trade-off between squeezing and purity which is illustrated in Fig.~\ref{fig:smaxsqpurity}. 

Interestingly, even the ubiquitous coherent state contains generalized superpositions with a distance of half a shot noise unit~\cite{Cavalcanti}. As the statistic of a coherent state is not altered by attenuation, these superpositions are immune to loss.

\begin{figure}
\includegraphics[scale=0.5]{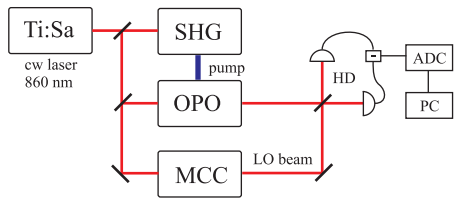}

\caption{\label{fig:exp-setup}Experimental setup for measuring a quadrature probability
distribution of squeezed states.
SHG: second harmonic generator,
MCC: mode cleaning cavity,
LO: local oscillator,
HD: balanced homodyne detecter.
}
\end{figure}

We now proceed with the experiment proving macroscopic coherence of Gaussian squeezed states sketched in Fig.~\ref{fig:exp-setup}.
For the generation of squeezed states we used a periodically poled
KTiOPO$_4$ (PPKTP) optical parametric oscillator (OPO)~\cite{PPKTPAPL}. The OPO was pumped by the second harmonic of a cw Ti:Sa laser (Coherent MBR110) at 430~nm, and the oscillation threshold was 180~mW. 
The squeezed states generated at 860~nm were measured using a balanced homodyne detector (HD). To ensure a high spatial overlap between the local oscillator and the squeezed states, the former was spatially cleaned using an empty cavity with a configuration identical that of the OPO. The output signal of the homodyne detector was digitized at a sideband frequency of 1~MHz with a resolution bandwidth of 30~kHz using an analog-to-digital converter (ADC, NI PXI-5124) and subsequently fed into a computer.
The total quantum efficiency was between 93.6\% and 94.4\% depending on the pump powers of the OPO. We measured squeezing between 3.7~dB and 7.7~dB associated with different pump powers of the OPO, and anti squeezing between 3.9~dB and 11.3~dB, respectively.

To prove the presence of macroscopic superposition states we record time series of the quadrature distributions of conjugate quadratures ($x$ and $p$) in separate runs. Subsequently we compute the variance of $p$ as well as the variances and mean values for the distinct regions resulting after binning the outcomes of the $x$ measurements.

First we experimentally demonstrated the proof of generalized superpositions
with distances of half a shot noise unit for vacuum states. The vacuum state was measured by blocking the input beam of the homodyne system and measuring conjugate quadratures as mentioned above. After calculating the relevant variances and using ineq. (\ref{eq:reid-unequality}), generalized superposition states were proven with a distance of $S=0.51$ SNU.

\begin{figure}
\includegraphics[scale=0.9]{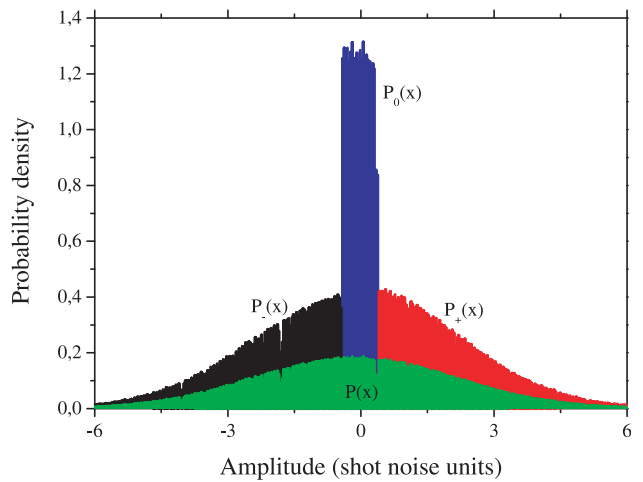}

\caption{\label{fig:probdist}Quadrature $x$ probability density distributions of binned regions and unbinned data of a squeezed beam (30~mW pump power) with distance $S=0.83\pm0.02$~SNU. The subspace probability distributions are each separately normalized.}
\end{figure}

To increase the maximum S, for which macroscopic superposition states can be witnessed, the
probability distribution of a measured quadrature has to become broader,
leading to a decrease of the variance of the conjugate variable. This can, as mentioned above, be accomplished with squeezed states. However,
often considerable amount of squeezing is accompanied with a loss
of pureness in real experiments, thus creating a trade-off between squeezing and purity.

For the squeezed states we proved
generalized superpositions with a maximum distance of $0.83\pm0.02$ shot noise units at 30 mW pump power, which
is significantly above the limit what can be proven with a coherent state. In Fig. \ref{fig:probdist} the probability density distribution $P(x)$ of the measurement values in the antisqueezing direction is plotted. The distributions $P_{-}(x)$ and $P_{+}(x)$ for the
$I=-1$ and $I=+1$ regions are separated by the distance $S=0.83\pm0.02$ SNU.

\begin{figure}
\includegraphics[scale=0.89]{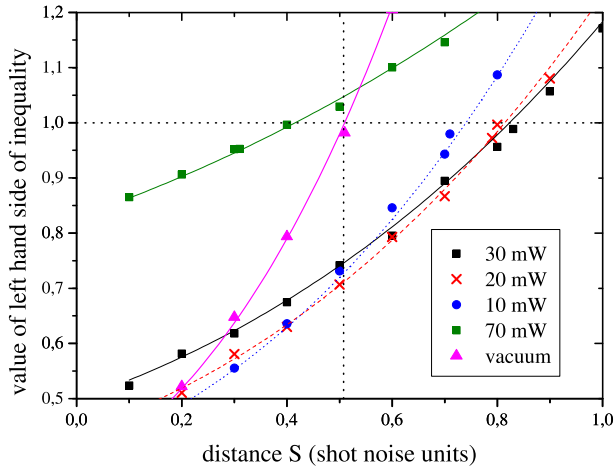}

\caption{\label{fig:squeezeuneq}Value of the left hand side of inequality (\ref{eq:reid-unequality}) vs. distance $S$ for different squeezed states and the vacuum coherent state. A value smaller than one proves the existence of a generalized superposition of distance $S$.}

\end{figure}

We studied the data with respect to the distance $S$ and different squeezing and purity levels. In Fig. \ref{fig:squeezeuneq} we show the value of the left hand side of inequality (\ref{eq:reid-unequality}) vs. the distance $S$.
Values calculated from the measurement data are depicted as symbols. The solid and dashed lines denote theoretical calculations
of the behavior of gaussian squeezed states with the measured variances.
A value less than one (indicated by the horizontal dotted line) proves the existence of generalized superposition with this distance. The distance $S\approx0.5$
marks the border of what is achievable with a coherent state. The behavior is investigated for the vacuum coherent state and different squeezed
states, that were generated by different pump powers of the OPO. As the pump power is increased more squeezing
is obtained. At the same time excess noise increases and the purity of the states drops. The best tradeoff
between squeezing and purity is found for a pump power of 30 mW, which results in squeezed states of -5.7~dB squeezing
and a purity of 0.85.
The purity shows its importance especially at the state of highest squeezing (pump power of 70~mW). Although -7.7~dB of
squeezing is measured, generalized superpositions were only proven with a maximum distance of $S=0.40\pm0.02$ SNU, because the purity
of the state dropped to 0.66. The vacuum coherent state shows a steep slope because of lack of squeezing and raises above 1 in inequality (\ref{eq:reid-unequality}) for distances larger than 0.51 SNU. In Fig. \ref{fig:smaxsqpurity} we plot the maximum distance ($S_{max}$) achieved for each pump power.

We also studied two other cases. In the first one two entangled beams were generated by two OPOs using a setup described in \cite{td-EPR}. In the second one we studied coherent states emitted from a low noise laser, such as the one used in \cite{low-noise}. We demonstrated the violation of inequality (\ref{eq:reid-unequality}) in both cases with a maximum distance of, respectively $0.30\pm0.02$ and $0.51\pm0.02$ SNU. For the entangled beams the S value was limited by the impurity of the states used. For the coherent beam with $10^{10}$ photons per measurement time interval we reproduced the result obtained for the vacuum state in Fig. \ref{fig:squeezeuneq}. Details will be reported elsewhere.

It is intriguing that in the ubiquitous coherent states sizeable generalized superpositions
can be proved. They serve as a
signature of the quantumness of a coherent state, regardless of its displacement in phase space. 
For several quantum states we proved the existence of generalized superpositions
with distances between the $\Psi_{+}$ and $\Psi_{-}$ regions approaching
one shot noise unit, which is comparable to the distances obtained in recent efforts on generating a "Kitten" state~\cite{kitten}.
Finding a generally applicable definition of macroscopicality of such superpositions
is not an easy task. The question is much debated also for other systems in quantum optics \cite{reviewleggett}. It is often not the overall number of photons in a mode which indicates macroscopicality but rather the effective number required to create the non classical nature of the field \cite{sidebandsq}. In a number of special cases measures of macroscopicality were suggested \cite{macromeasure}. A connection to the definition of
generalized superpositions as defined in \cite{Cavalcanti} will be the goal of future investigations.

%\begin{acknowledgments}
We like to thank E.G. Cavalcanti, M.D. Reid, N. Treps and C. Fabre for useful discussions. GL acknowledges support from the EU project COVAQIAL (FP6-511004). AF acknowledges financial support from Japan Science and Technology 
Agency and the MEXT of Japan.
%%%%%
%\end{acknowledgments}

\end{document}